\documentclass[superscriptaddress,twocolumn,10pt]{revtex4-1}

\usepackage{amsmath,graphicx,color}
\usepackage{booktabs,array,dcolumn}

\usepackage{multirow}
\usepackage{times}
\usepackage{xcolor,soul}
\newcolumntype{C}{>{\centering}p}
\renewcommand{\thefootnote}{\fnsymbol{footnote}}
\newcommand{\etal}{\emph{et al.}}

\begin{document}

\title{Phase transitions and cluster structures of the new finite range Lennard-Jones like model}

\author{Omar-Farouk Adesida}
\altaffiliation{Contributed equally to this work}
\affiliation{Department of Physics, University of Warwick, Coventry, CV4 7AL, UK}

\author{Sebastian Havens}
\altaffiliation{Contributed equally to this work}
\affiliation{Department of Chemistry, University of Warwick, Coventry, CV4 7AL, UK}

\author{Livia B. P\'artay}
\email{Livia.Bartok-Partay@warwick.ac.uk}
\affiliation{Department of Chemistry, University of Warwick, Coventry, CV4 7AL, UK}

\date{\today}

\begin{abstract}
In the current work we revisit the pair-potential recently proposed by Wang \etal{} (\emph{Phys. Chem. Chem. Phys.} 10624, 22, 2020) as a well defined finite-range alternative to the widely used Lennard-Jones interaction model. 
The advantage of their proposed potential is that it not only goes smoothly to zero at the cutoff distance, hence eliminating inconsistencies caused by different treatments of the truncation, but with changing the range of the potential, it is capable of describing soft matter-like behaviour as well as traditional ``Lennard-Jones-like'' properties. 
We used the nested sampling method to perform an unbiased sampling of the potential energy surface, and mapped the pressure-temperature phase diagram of a range of truncation distances. We found that the interplay between the location of the energy minimum and interaction range has a complex and strong effect on both the structural and thermodynamic properties of the condensed phases.
We discuss the appearance of the liquid-vapour co-existence line and critical point at longer interaction ranges, as well as the relatively small changes in the melting line.
We present the ground state diagram, demonstrating that different close-packed polytypic phases appear to be global minima for the model at different pressure and cutoff values, similar to what had been shown in case of the Lennard-Jones model. Finally, we compare the lowest energy structure of some of the $N<60$ clusters to that of the known minima of Lennard-Jones and Morse potential, using different cutoff values, revealing a behaviour closely resembling that of the Morse clusters.

\end{abstract}

    \maketitle

\section{Introduction}

Large-scale simulations and computational method development often relies on the use of well known and simple pair potentials that are computationally cheap to evaluate. One of the most widely used such potentials is the so-called Lennard-Jones (LJ) 12-6 model,
\begin{equation}
    \phi(r)=4\epsilon\Bigg( \Big[\frac{\sigma}{r} \Big]^{12} -  \Big[\frac{\sigma}{r} \Big]^6\Bigg), 
\end{equation}
proposed almost a hundred years ago by John Lennard-Jones \cite{LJ1,LJ2}
to describe the cohesive energy between noble gas atoms. Since then it has become the most widely employed potential function in classical simulations. It is frequently used as a fictitious model material, Lennard-Jonesium, acting as a test system for method development and bench-marking simulations, while also forming the bases of molecular interaction models, coarse-grain potentials\cite{LJ_coarse, LJ_coars2} and even used to describe interactions in cellular systems\cite{cell_LJ2,cell_LJ1}.  
Despite all the computational advantages, appealing general properties and extensive usage in computer simulations, the LJ 12-6 model suffers from some significant drawbacks, summarised by Wang and co-workers in their recent work\cite{WangLJ}.
Most importantly, the LJ potential has an infinite range that requires truncation for practical computational purposes. 
The choice of this cutoff distance and further optional modifications to control the potential's approach to zero at the cutoff (e.g. energy-shifted or force-shifted versions of the model) can significantly change its behaviour. 
For example, different truncation distances and procedures result in widely different surface tension values,\cite{LJ_surf_tension} different critical properties,\cite{smit1992phase} solid-fluid coexistence\cite{LJ_melting} and even affect the relative stability of crystalline phases and the ground state structure.\cite{LJPolytypism,Ackland_stacking,LJ_solid} 
These differences in structural and thermodynamic properties highlight the fact that the LJ 12-6 potential is not unambiguously defined when truncated in practical applications, and hence not the most practical choice as a standard reference system.
Moreover, the widely used 12-6 exponents do not reflect the shorter range interactions of e.g. nano-colloids, and hence do not provide the necessary flexibility to capture the properties of larger than atomic-scale systems.

To overcome these shortcomings, Wang \etal{} proposed a new family of potential models that are well defined, approaching zero smoothly and continuously at the cutoff. Moreover, the potential is able to capture the behaviour of a wide range of materials classes by simply changing a single interaction range parameter.\cite{WangLJ}  
In their work they evaluated the bulk density-temperature phase diagram and thermodynamic properties of two separate model parameters ($r_c=1.2\sigma$ and $r_c=2.0\sigma$), using various traditional simulation techniques. 

For this better defined Lennard-Jones-like model to become a truly useful reference system for simulations, a more detailed understanding of its properties is necessary, and for a wider range of truncation parameters. Since its original publication, the thermal conductivity of $r_c=2.0\sigma$ has been studied\cite{cheng2020}, the fluid behaviour of the short range colloid-like system of $r_c=1.2\sigma$ described in detail\cite{mausbach2022} as well as the face-centred-cubic and hexagonal-close-packed phase boundaries determined for interaction ranges from $r_c=2.0\sigma$ to $r_c=7.0\sigma$\cite{moro2024}.

In the current work we wish to extend these descriptions and contribute to the more detailed general understanding of the behaviour of the model. We calculated the pressure-temperature phase diagram of multiple further cutoff parameters with the aid of the nested sampling method\cite{NS_all_review} in order to investigate the change from soft matter like to Lennard-Jones like behaviour. We systematically investigated the dependence of the ground state structure on the pressure and the interaction range. Finally, we also studied the global minimum structure of up to medium sized clusters, in order to compare them to those of the Lennard-Jones model.


\section{Computational details}

\subsection{The WRDF potential model}

\begin{figure}[hbt]
\begin{center}
\includegraphics[width=6.5cm,angle=90]{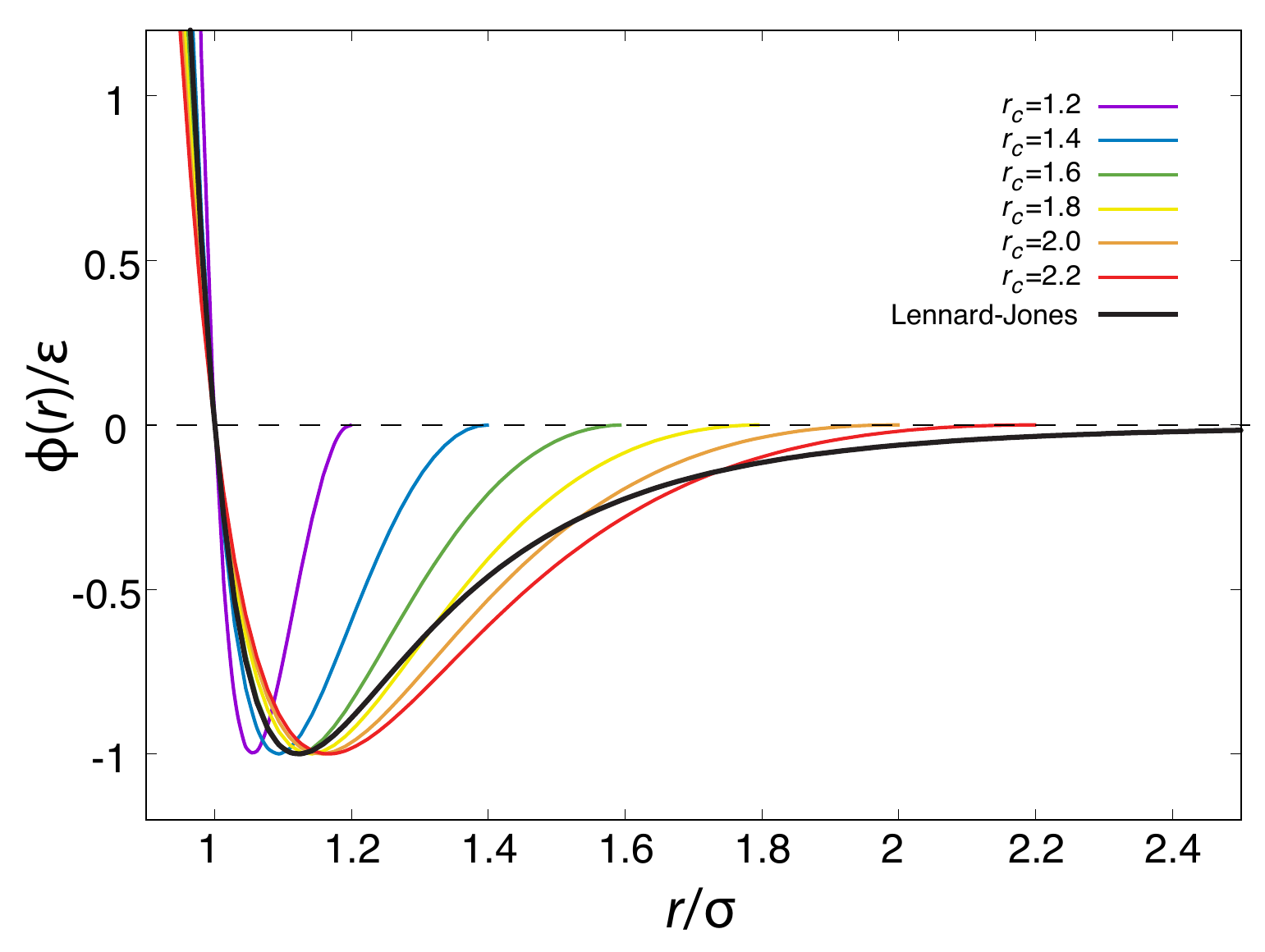}
\end{center}
\vspace{-20pt}
\caption {Comparison of the WRDF potential functions with different $r_c$ values and with $\mu=\nu=1$ to the original un-truncated 12-6 Lennard-Jones potential.} 
\label{fig:WF_pot}
\end{figure}

The general functional form proposed by Wang \etal{}\cite{WangLJ} is 
\begin{equation}
    \phi(r)=\epsilon\alpha\Bigg( \Big[\frac{\sigma}{r}\Big]^{2\mu}-1\Bigg)\Bigg( \Big[\frac{r_c}{r}\Big]^{2\mu}-1\Bigg)^{2\nu}, 
\end{equation}
where $r_c$ is the cutoff distance where the potential vanishes quadratically (i.e. both the energy and the first derivative is zero at $r_c$, but the second derivative has a finite value), and $\mu$ and $\nu$ are positive integers. Parameters $\sigma$ and $\epsilon$ act as controlling parameters for distance and energy, respectively, similar to the original Lennard-Jones model. 
Parameter $\alpha$ ensures that the minimum energy of the potential is $-\epsilon$ and hence it is
\begin{equation}
    \alpha(\nu,\mu, r_c)=2\nu r_c^{2\mu} \Bigg(\frac{1+2\nu}{2\nu\big(r_c^{2\mu}-1\big)}\Bigg)^{1+2\nu}.
\end{equation}
We will refer to this potential as the WRDF model, crediting the authors of the original work. Figure~\ref{fig:WF_pot} shows the shape of the potential using a range of $r_c$ values studied in the current work, compared to the shape of the un-truncated 12-6 Lennard-Jones potential.

\subsection{Computational details}

We used the nested sampling (NS) algorithm to perform the unbiased sampling of the configuration space of the model. NS was originally introduced in the area of Bayesian statistics\cite{bib:skilling,bib:skilling2,NS_all_review} and has been adapted to sample the potential energy surface of atomistic systems\cite{1st_NS_paper, NS_mat_review}, allowing the automatic generation of thermodynamically relevant configurations from the gas phase to the ground state structure, without any prior knowledge of e.g. crystalline phases or the location of phase transitions, making it a fully predictive algorithm. 
NS provides a unique and easy access to the partition function at an arbitrary temperature during the post-processing of the generated samples. 
By locating peaks in the second derivatives of the partition function, such as the heat capacity or the compressibility, we can easily locate phase transitions. This means that NS allows us to draw a map of the entire pressure-temperature phase diagram unbiased, allowing the discovery of phase transitions as well as solid phases, using a single computational technique. 
The power of NS has been demonstrated in studying the phase behaviour of various systems, including clusters\cite{1st_NS_paper,CuPt_ns,dorrell_thermodynamics_2019,NS_wales}, metals\cite{pt_phase_dias_ns,ConPresNS,NS_lithium}, alloys\cite{CuAu_Pastewka,AgPd_ML} and model potentials\cite{NS_hardsphere,jagla,jagla_2D}. 
While the computational cost of NS requires the use of relatively small systems sizes, resulting in e.g. the slight overestimation of the melting transition,\cite{NS_mat_review} these values can be easily refined with other specialised methods if a higher accuracy is necessary. We demonstrated in previous works, that using the structures and transition points identified by NS as a guide for such refinements, is often critical to be able to elucidate the true behaviour of a potential, highlighting the real power of NS.\cite{jagla,NS_lithium} 

The nested sampling calculations in the current work were performed as presented in Ref.~\cite{pt_phase_dias_ns}. 
Initial sample configurations were generated by placing the particles randomly in the simulation cell of volume $500\sigma^3/\mathrm{atom}$. New samples were generated by performing 1000 Monte Carlo steps, including particle moves and changing the volume and shape of the simulation cell by shear and stretch moves, with a step probability ratio of 20:6:3:3, respectively. 
The simulations were run at constant pressure in the range of $p=(0.01-50)~\epsilon\sigma^{-3}$, and the bounding cell of variable shape and size contained 32 particles, or 64 for more accurate sampling of the vapourisation transition. 
While these system sizes appear to be small, they allow the systematic exploration of the configuration space of a large number of different pressure and potential parameter values, capturing thermodynamically relevant phases.\cite{NS_mat_review}
The number of walkers, $K$, controls the resolution of the sampling, with the computational cost depending on it linearly. $K$ is usually chosen such that the resulting heat capacity peaks are sufficiently converged. In our case the range $K=528-1008$ was found to be sufficient to achieve less than 3\% variance in the calculated melting temperature (see variance between parallel nested sampling runs on Figure~\ref{fig:P1_compare}). 

At the end of the sampling, the isobaric partition function, $\Delta$, can be computed as 
\begin{equation}
    \Delta(N,\beta,p) \approx \sum_i (\Gamma_{i-1}-\Gamma_{i}) e^{-\beta (E_i+pV_i)},    
\label{eq:average}
\end{equation}
where $N$ is the number of particles, $\beta$ is the inverse temperature; $E_i$, $V_i$ and $\Gamma_i$ are the energy, volume and phase space volume of the configuration recorded in the $i$-th NS iteration respectively, with $\Gamma_i$ calculated as $\Gamma_i=(K/(K+1))^i$.\cite{pt_phase_dias_ns}
We can also calculate the phase space weighted average of observables (e.g. density or the radial distribution function) to evaluate their finite temperature values using the following equation:
\begin{equation}
    \langle A \rangle \approx \frac{1}{\Delta}\sum_i A_i (\Gamma_{i-1}-\Gamma_{i}) e^{-\beta H_i},    
\label{eq:average}
\end{equation}
where $A_i$ is the observable value recorded in the $i$-th NS iteration.\cite{1st_NS_paper}

Nested sampling for small clusters were performed similarly to the bulk simulations, starting from random configurations, but in a cubic box of fixed volume of $50\sigma^3$ per particle, and using $K=1008$ walkers.

A parallel implementation of the NS algorithm is available in the {\tt pymatnest} python software
package\cite{pymatnest}. To calculate the energy and propagate the configurations during the sampling, {\tt pymatnest} can be used either with a Fortran implementation of the WRDF potential or with the \texttt{wf/cut} pairstyle in the LAMMPS package\cite{LAMMPS}.

We performed a series of structure optimisations, both for bulk phase crystal structures as well as cluster configurations. These minimisations were performed using the LAMMPS package\cite{LAMMPS} with \texttt{pairstyle wf/cut}, with energy and force tolerances of $10^{-6}\epsilon$ and $10^{-8}\epsilon\sigma^{-1}$, respectively. 

Basin hopping simulations of clusters were performed via our implementation employing ASE, using the methodology prescribed in Ref. \cite{bib:wales_globopt_LJ}, starting from a random atomic configuration. Basin hopping was carried out using 20000 steps, with stepsizes adjusted such that the acceptance ratio of translational and angular moves (i.e. moving a particle around the centre of mass of the cluster) is between 0.2 and 0.5.

\section{Results}

\subsection{The phase diagram}

The WRDF model does not only alleviate the ambiguity arising from the truncation of the LJ model, but it also offers more varied and tuneable phase behaviour. 
Understanding these, for example phase transition properties or how the colloid-like behaviour changes to atomic-like behaviour with increasing cutoff values is essential if the model is to be adapted in a wide range of classical simulations. 
The exhaustive sampling of its potential energy surface can also give a valuable insight into how the macroscopic properties of the model system depend on details of the pair interaction. 

\begin{figure}[hbt]
\begin{center}
\includegraphics[width=8.5cm,angle=0]{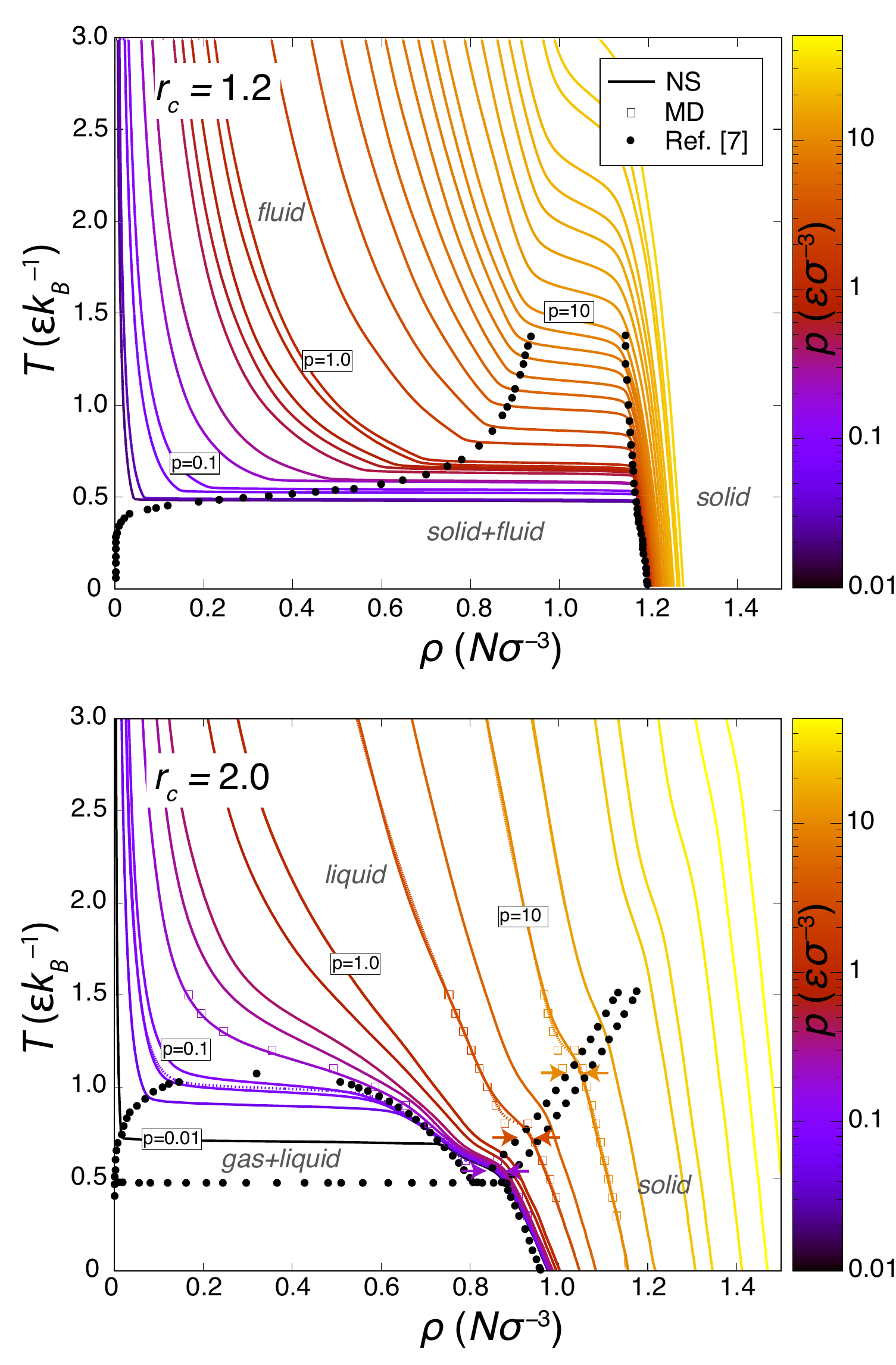}
\end{center}
\vspace{-20pt}
\caption {Density-temperature phase diagram, calculated using cutoff parameters 1.2 (top panel) and 2.0 (bottom panel). Lines are isobars sampled by nested sampling (NS), coloured according to the corresponding pressure. Black circles mark phase transitions from Ref.~\cite{WangLJ}. In the bottom panel, open squares mark densities of the solid and fluid phases at different pressures, sampled by MD simulations. Arrows mark transitions determined by two-phase coexistence simulations.} 
\label{fig:rhoT_phaseD}
\end{figure}

The original work of Wang \etal{}\cite{WangLJ} presented the density-temperature phase diagrams of the WRDF potential at two cutoff values, $r_c=1.2\sigma$ and $r_c=2.0\sigma$.
To verify our nested sampling calculations, we first compared the transition points obtained from density-temperature isobars at different pressures to the results of Wang \etal{}\cite{WangLJ} in Figure~\ref{fig:rhoT_phaseD}.
The nearly horizontal sections of the isobars correspond to regions where two phases (e.g. solid-fluid) co-exist, and sharper changes in the gradient correspond to phase transitions. 
In case of $r_c=2.0\sigma$ we also performed a series of molecular dynamic simulations of 864 particles using the $NpT$ ensemble, to compare the solid and liquid densities (shown by open squares in Fig.~\ref{fig:rhoT_phaseD}) with those calculated by NS. 
These results show very good agreement with the phase diagrams published in Ref.~\cite{WangLJ}, demonstrating that NS samples the same phase transitions. 
We also performed two-phase coexistence simulations at three different pressures for $r_c=2.0\sigma$ (highlighted by arrows in Fig.~\ref{fig:rhoT_phaseD}), to be able to estimate the finite size effect of the small system size sampling. 
In agreement with previous works\cite{NS_mat_review}, we found that the fluid-solid phase boundary is reliably reproduced with even only 32 particles, with the transition point being overestimated by approximately $8-10\%$. 

\begin{figure}[hbt]
\begin{center}
\includegraphics[width=8.5cm,angle=0]{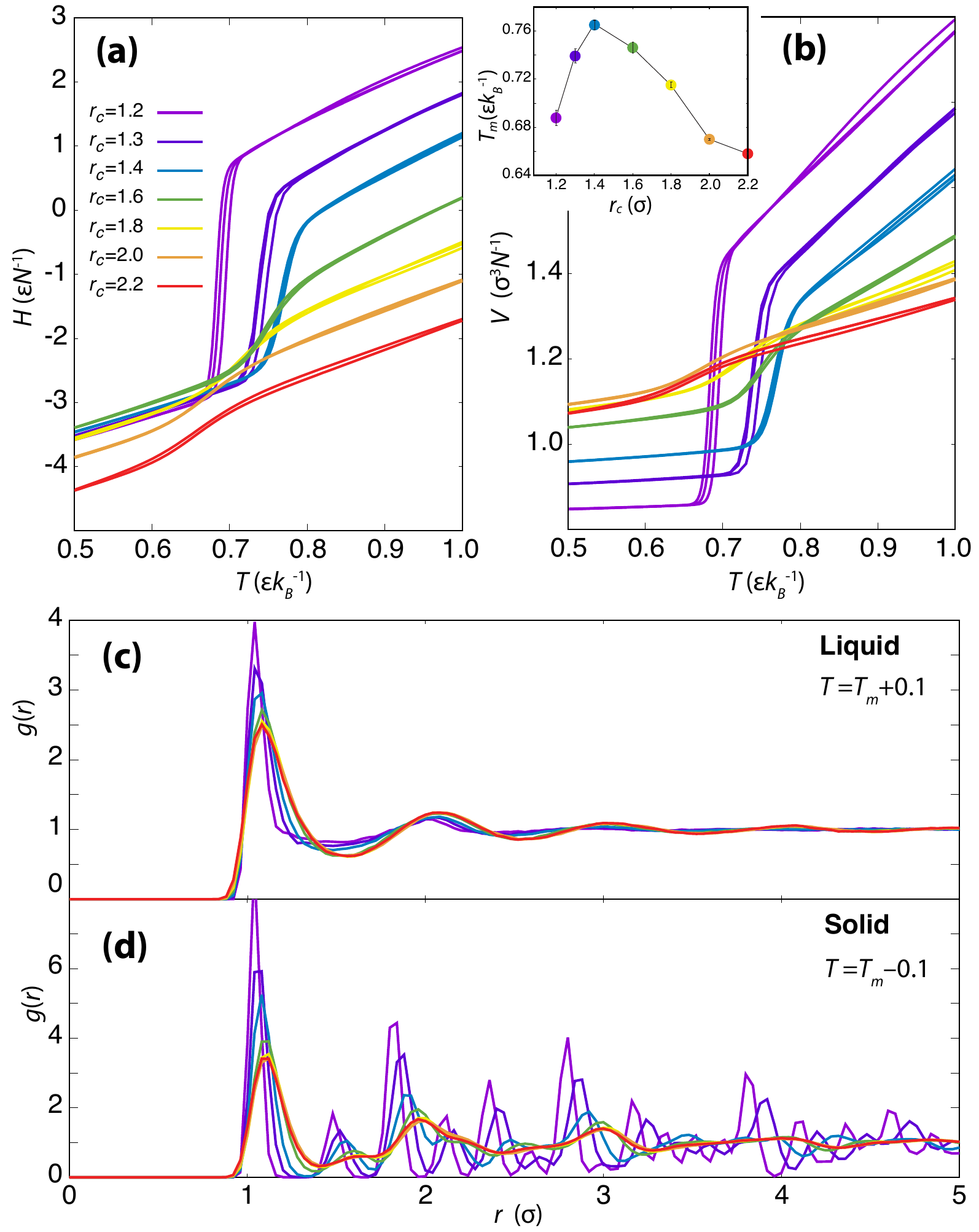}
\end{center}
\vspace{-20pt}
\caption {Enthalpy (panel (a)) and volume (panel (b)) as a function of temperature around the melting transition from three parallel nested sampling calculations at each different spatial cutoff values, using 32 particles at $p=1.0~\epsilon\sigma^{-3}$. The inset shows the melting temperature as a function of the cutoff distance at this pressure. Panels (c) and (d) show the radial distribution functions of the liquid and solid phases at $0.1\epsilon k_B^{-1}$ above and below the melting temperature, respectively.} 
\label{fig:P1_compare}
\end{figure}

For WRDF it is important to consider that -- in contrast to the LJ model -- not only the range of the interaction changes with changing the value of $r_c$, but also the shape and the location of the energy minimum (as illustrated in Figure~\ref{fig:WF_pot}). Therefore, the density and the energy of condensed phases, and hence the latent heat of transitions, will be influenced by the interplay between these two affects.
Figure~\ref{fig:P1_compare} demonstrates this, showing the enthalpy and volume changes around the melting transition, comparing the effect of various potential cutoff values at $p=1.0~\epsilon\sigma^{-3}$.
It is apparent that the enthalpy of the liquid phase becomes more negative with increasing $r_c$, due to an increasing number of neighbours falling within the cutoff radius, while
the enthalpy of the solid phase is much less sensitive to changes in the potential truncation. In a close-packed crystal structure only the first neighbour shell contributes to the energy if $r_c<1.8\sigma$, therefore the energy of the solid phase does not depend at all on the value of $r_c$ below this limit. These opposing trends mean that the latent heat of the melting transition becomes smaller for longer range interactions.
As expected for a LJ-like pair potential, the solid phase of WRDF particles is more dense than its liquid phase, thus the volume increases upon melting. However, this volume increase is also strongly affected by the choice of the cutoff parameter: as the value of $r_c$ (and thus the minimum energy distance) increases, the volume of the solid phase increases, while the truncation distance has an opposite effect on liquid density.
Considering entropic effects can help us understand these trends: at very short interaction ranges entropic effects are larger than the energy contribution of a very narrow neighbour shell, while at longer range interactions energetic contributions will have more influence. This can also be seen on the average radial distribution functions, calculated $0.1\epsilon k_B$ below and above the melting temperature (see panels (c) and (d) in Fig.~\ref{fig:P1_compare}, respectively). At small $r_c$ values the solid phase is significantly more ordered, while at larger values the liquid phase appears to be more structured, thus causing the entropy difference between the two phases to become smaller as the potential truncation is increased.  

\begin{figure}[hbt]
\begin{center}
\includegraphics[width=8.5cm,angle=0]{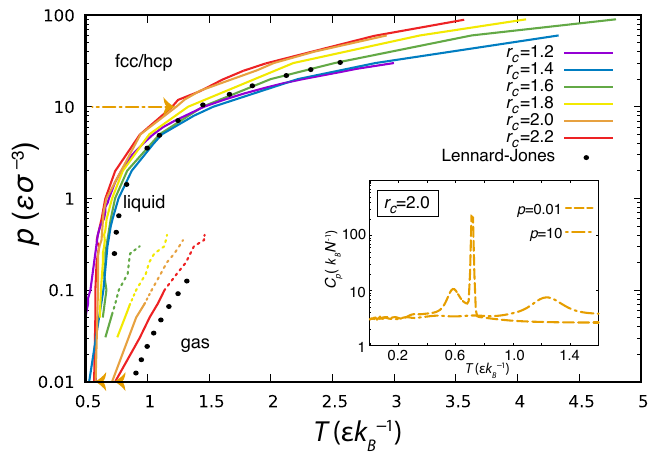}
\end{center}
\vspace{-20pt}
\caption {Pressure-temperature phase diagram, calculated using different cutoff parameters. Dashed lines correspond to the Widom-line of the heat capacity, above the liquid-vapour critical point. The inset shows two heat capacity curves used for the identification of transition points, as examples, for $r_c=2.0\sigma$, at two different pressures, which are also marked by arrows on the main plot. Black symbols show transitions points of the Lennard-Jones system for comparison, with melting temperature data taken from Ref.~\cite{mcneil2006freezing} and boiling point data from Ref.~\cite{agrawal95}.} 
\label{fig:PT_phaseD}
\end{figure}

Interestingly, not only the thermodynamical characteristics of the condensed phases change, but also the location of the transition. For shorter range potentials, the melting temperature increases up to $r_c=1.4\sigma$, then decreases as the cutoff value increases further (see inset of panel (b) in Fig.~\ref{fig:P1_compare}). 
However, as pressure has a large influence on the volume and thus the enthalpy of the phases, this trend changes with pressure. 
Figure~\ref{fig:PT_phaseD} shows the pressure-temperature phase diagram, calculated using a range of different cutoff values.
The overall trends of the melting lines follow the generally expected behaviour: the melting temperature increases with increasing pressure, and this increase is more significant at higher pressures.
The effect of the interaction's range is weak, but the slope of the melting line appears to increase for larger cutoff values. This is most pronounced for the potentials displaying colloid-like behaviour, $r_c=1.2\sigma$ and  $r_c=1.4\sigma$, which intersect the other melting lines.  

A more obvious effect of increasing the interaction range is the appearance and position of the liquid-gas phase boundary on the phase diagram. 
While sampling the sublimation curve directly is challenging, due to the large entropy difference between the gas and crystalline phases, the trends in the location and slope of boiling curves can help us estimate at which $r_c$ value the liquid and gas phase becomes distinguishable. 
For $r_c\geq1.4\sigma$ values we found a single heat capacity peak, even at low pressure, corresponding to the solid-fluid transition, but at $r_c=1.6\sigma$ two peaks appear close to each other in temperature. One transition has a relatively small density change (solid-liquid transition), while the higher temperature transition causes a significant change in the volume (liquid-gas transition). 
With changing the value of $r_c$ we found that the boiling temperature at the same pressure increases for longer interaction ranges, however, the gradient of the boiling curve does not change significantly.  
Extrapolation from these findings suggest that the triple point is likely to appear in the cutoff range $1.4\sigma \le r_c \le 1.6\sigma$, beyond which the triple point pressure rapidly decreases as the cutoff is increased further. 

In case of the gas-liquid phase boundary, a peak on the thermodynamic response functions, such as the isobaric heat capacity or compressibility, can still be observed in the supercritical region, corresponding to the Widom-lines.\cite{xu2005relation,simeoni2010widom}
Though such peaks get shallower and broader rapidly beyond the critical point, it is difficult to locate the critical point using only the shape or characteristics of the peak in our case. Since we sample finite size systems with NS, the heat capacity does not diverge at first order phase transitions as expected in the thermodynamic limit, but a finite and broader peak can be observed. To be able to differentiate peaks corresponding to first-order gas-liquid transitions from those corresponding to the Widom-line, and estimate the lower and upper bounds of the critical pressure, we calculated the volume distribution of the configurations generated by NS around the transition -- a bimodal distribution signals a first-order gas-liquid transition, while a unimodal distribution means a continuous transition, thus being above the critical pressure.\cite{bruce1992scaling} 
Our results suggest that both the critical temperature and pressure increases with the interaction range. 
Widom-lines usually follow different paths on the pressure-temperature plane of the phase diagram, with the one corresponding to the isobaric heat capacity typically close to the critical isochore with a slight deviation towards higher densities.\cite{simeoni2010widom,brazhkin2011widom} 
In case of the Lennard-Jones potential, the heat capacity Widom-line diminishes at $T\approx 2.5T_c$ and $p\approx 10p_c$.\cite{brazhkin2011widom} 
Overall we can see a similar behaviour for the WRDF potential: above the critical point the heat capacity peaks shift towards higher densities than the critical value, but diminish at much lower pressure and temperature than for the Lennard-Jones potential.  

As expected for a Lennard-Jones-like model, the particles form close-packed crystalline structures below the melting point.
Depending on the cutoff, temperature and pressure, nested sampling identifies either the face-centred-cubic (fcc) or hexagonal-close-packed (hcp) structures, as well as other stacking variants closer to the solid-liquid transition boundary, especially for larger cutoff and pressure values.

While observable properties of the fcc and hcp phases of the WRDF model are very similar\cite{WangLJ}, previous works have shown that the truncation distance and smoothness of a potential and its derivatives at the cutoff can have a significant effect on the ground state structure.\cite{LJPolytypism, Ackland_stacking,moro2024} 
Hence, we performed a series of geometry optimisations in order to compare the stability of the following different close-packed stacking sequences: fcc, hcp, hc (also known as the double-hexagonal-close-packed structure), hhc (also known as the 9R structure), hhhc, hcc, hccc and hhcc. In this notation we mark a layer ‘‘c’’ as cubic, if it has fcc surroundings (the two neighbouring layers occupy different positions, e.g. \textbf{A}B\textbf{C}), while the layers with hcp surrounding (sandwiched between two layers occupying the same relative lateral position, e.g. \textbf{A}B\textbf{A}) is marked ‘‘h’’ as hexagonal.\cite{LJPolytypism}
The constant pressure geometry optimisations were performed using LAMMPS, fully relaxing both the internal coordinates and the lattice parameters. Starting configurations were created with atoms placed at $0.8\sigma$ distance from each other in the required stacking order, using a system size to allow the use of orthogonal lattices.
During the optimisation the stacking sequences did not change, the particles forming the stacking planes remained perfectly in plane and the lattice preserved its orthogonal symmetry.
After the optimisation, the enthalpies of the configurations were compared to identify the most stable structure.

\begin{figure}[hbt]
\begin{center}
\includegraphics[width=7.4cm,angle=0]{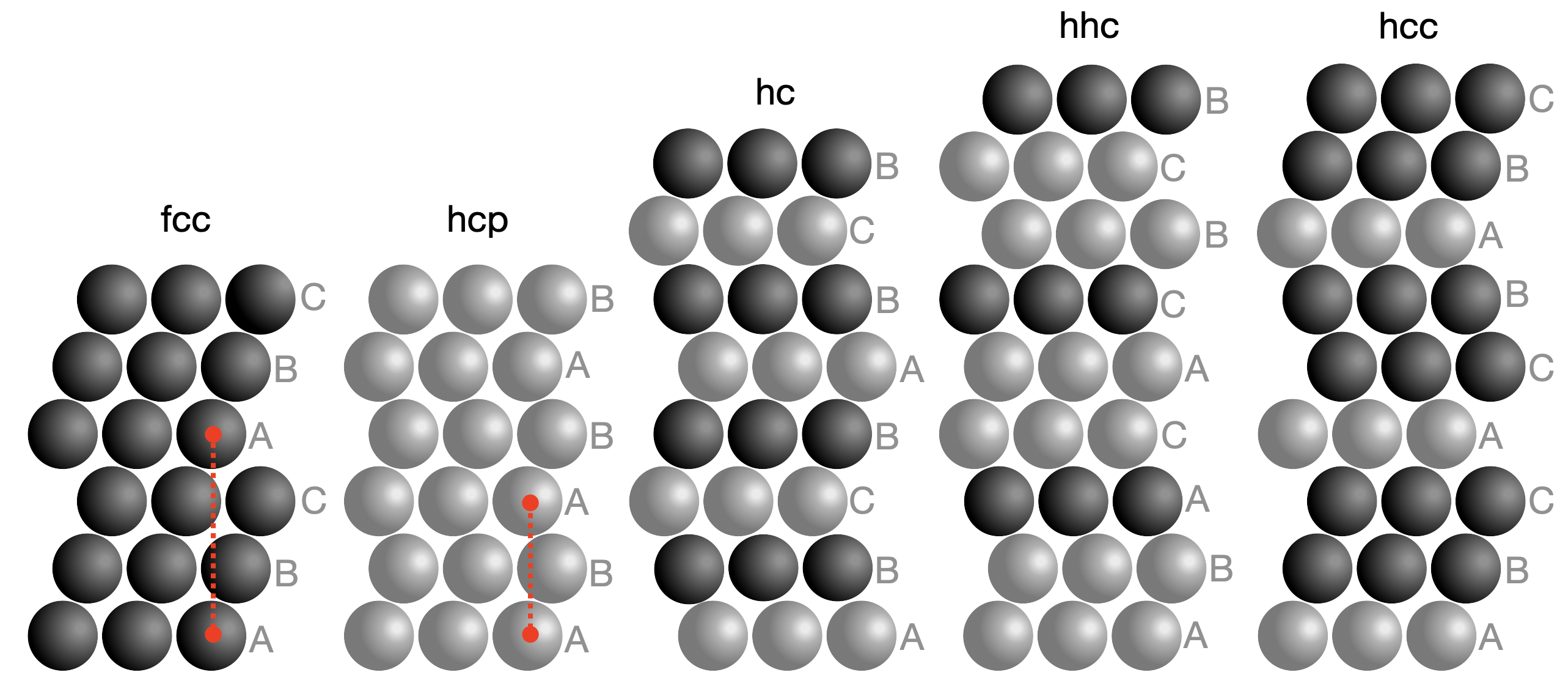}
\includegraphics[width=6.4cm,angle=90]{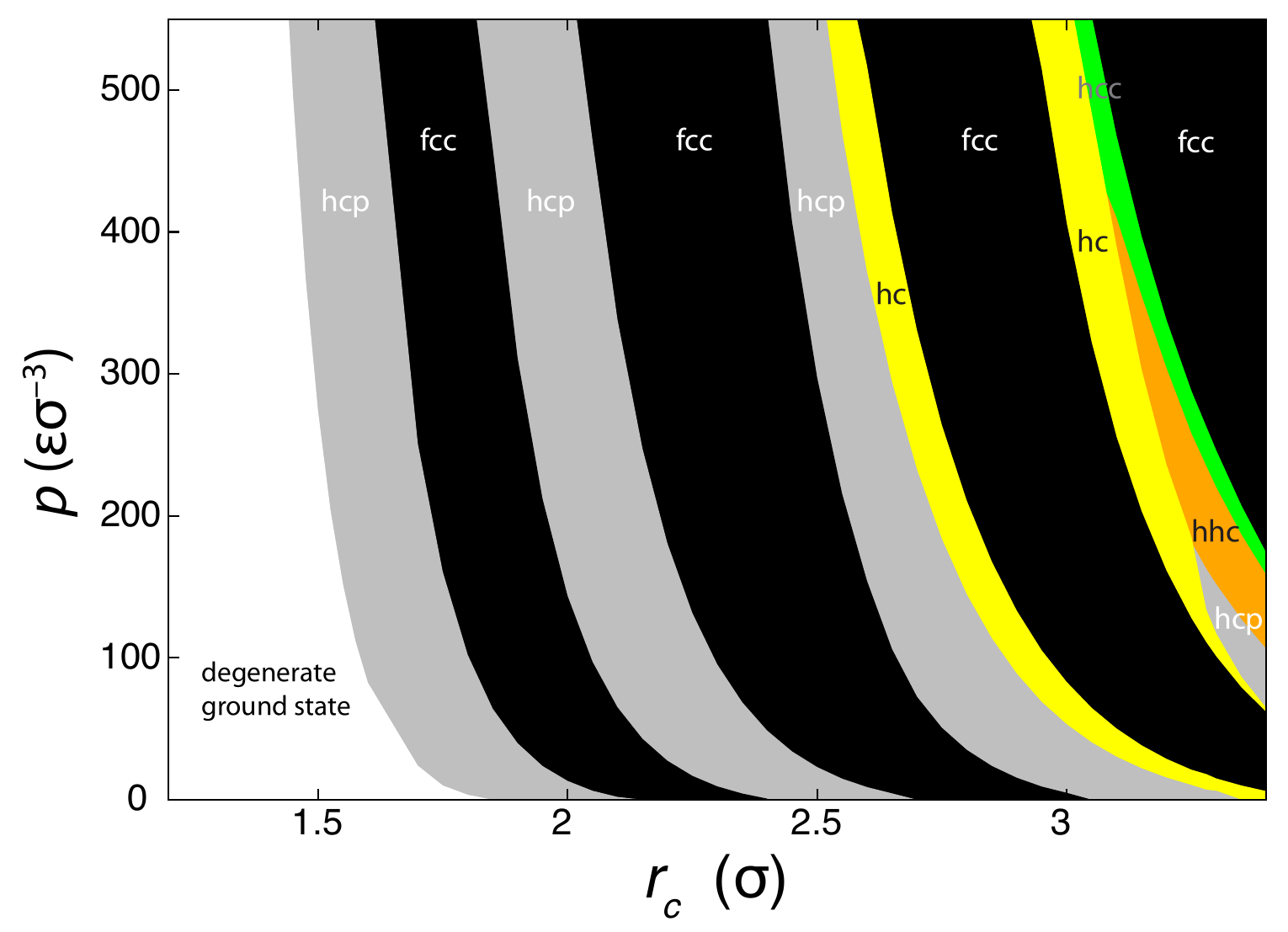}
\end{center}
\vspace{-20pt}
\caption {Ground state structure of the WRDF potential as a function of cutoff distance and at different pressures. The colours represent the regions where different stacking sequences are the global minimum: hcp (grey), fcc (black), hc (yellow), hhc (orange), hcc (green). The region where these structures are degenerate is left white. Stacking sequences of these structures are also demonstrated above the diagram: atoms with a cubic stacking environment are coloured black, while atoms with hexagonal stacking environment are coloured grey. The ABC notation is also shown for reference, with layers in the same stacking position highlighted with red lines for the fcc and hcp structures.} 
\label{fig:ground_state}
\end{figure}

The resulting ground-state phase diagram is shown in Figure~\ref{fig:ground_state}. 
At very short cutoff values and at lower pressures, only the 12 nearest neighbour particles are within the interaction range, and thus give contribution to the energy. Since this is identical for all different close-packed polytypic structures, their energies and enthalpies are degenerate. We marked this region white on the phase diagram.  
The overall ground-state behaviour shows similarities to that of the truncated Lennard-Jones potential.
Though the potential energy and the forces of the WRDF model approach zero smoothly at the cutoff distance, the second derivative of the potential function does not, and hence we can expect similar trends emerging as for the truncated and force-shifted LJ model.\cite{LJPolytypism} 
At $r_c>1.8\sigma$, with more and more neighbour shells entering the cutoff distance, regions of different stacking variants indeed appear as bands, with the hcp structure favoured towards lower pressures.
First, fcc and hcp alternate each other as the ground state structures, in agreement with the findings of Moro \etal{}\cite{moro2024}, with no other polytypes appearing to be the global minimum. 
However, as the number of particles within the cutoff range increases (e.g. at $r_c=3.0\sigma$ and $p=100 ~\epsilon\sigma^{-3}$ 320 particles are within the cutoff radius of the fcc structure) first the hc configuration, then also the hhc and hcc phases appear in regions as the most stable crystalline structure.
We note, that as shown by Loach and Ackland, the hhc and fcc phases have no direct phase boundary between them.\cite{Ackland_stacking} 

\subsection{Clusters}

Lennard-Jones clusters have long served as popular benchmark systems for testing the performance of sampling algorithms, and hence their structural and thermodynamic properties have been explored extensively and described in detail up to several hundreds of particles.
Magic numbers of particularly stable cluster sizes, low-temperature structural transitions, and the peculiar truncated octahedron fcc structure of 38 Lennard-Jones particles all attracted widespread interest both as general sampling challenges and computationally cheap model systems of nanoparticles. \cite{bib:energy_landscapes, bib:wales_basin_LJ, bib:wales_globopt_LJ, bib:wales_disconnectivity,1st_NS_paper,bib:vladimir,bib:M-aM_Northby}
The case of periodic systems highlighted that while at larger $r_c$ values the WRDF potential shows similar features and macroscopic properties as the Lennard-Jones potential, at shorter truncation distances the potential more closely resembles soft matter behaviour.
This naturally raises the question, whether the extensive knowledge of the LJ clusters and their special cases transfer to clusters described by the WRDF potential.  
In order to answer this question we optimised the known LJ and Morse global minima structures taken from The Cambridge Energy Landscape Database\cite{CCDB} using the WRDF potential at a range of different $r_c$ parameters.  
We also extended the search by performing nested sampling simulations and a series of Basin Hopping calculations\cite{bib:wales_basin_LJ} with different number of WRDF particles.
However, these calculations did not find any minima configurations that had lower energy than those already included in the database.
While one can never rule out the possibility of more extensive searches finding a new global minimum, we nevertheless can be confident that for the majority of cluster sizes the structures considered in the current work are representative.

\begin{figure}[hbt]
\begin{center}
\includegraphics[width=8.5cm,angle=0]{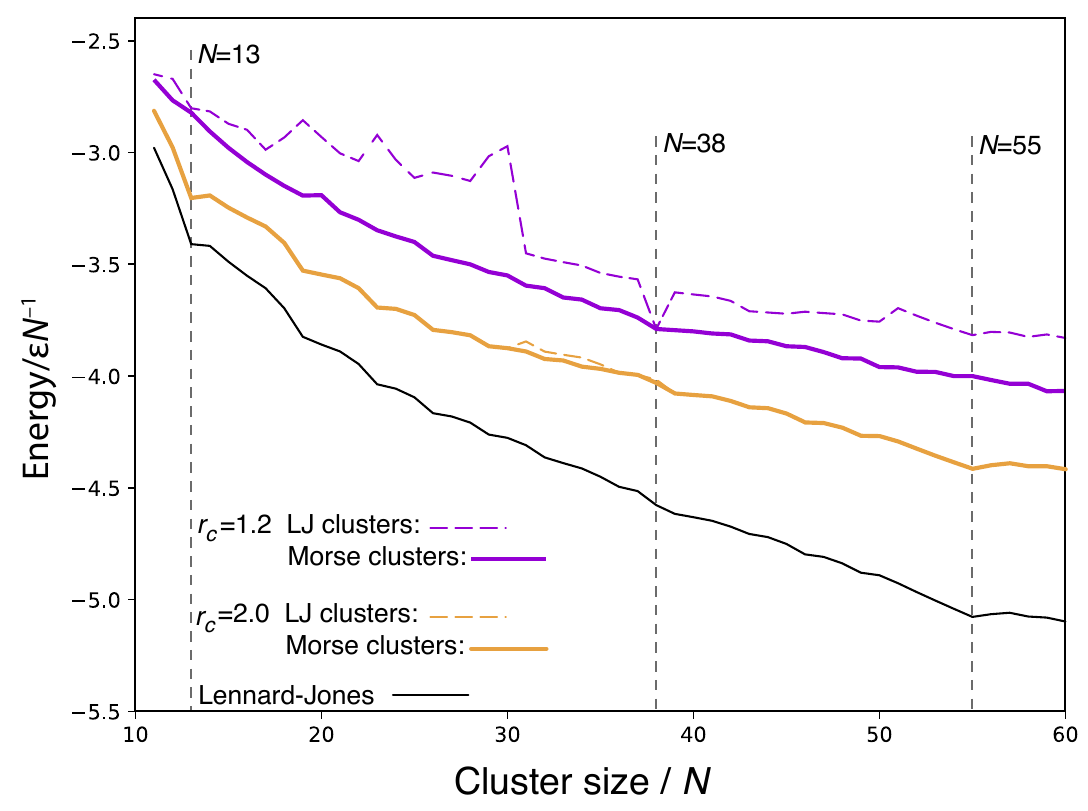}
\end{center}
\vspace{-20pt}
\caption {Energy per atom of the global minimum structure configuration, optimised using different $r_c$ values for the WRDF potential  (coloured lines) as well as using the Lennard-Jones potential with cutoff of $2.5\sigma$ (black line).
Dashed lines show the energies of structures starting the optimisation from known global minima of LJ clusters, solid coloured lines show the energy starting from known global minima of Morse clusters.} 
\label{fig:Cluster_LJ}
\end{figure}

Figure~\ref{fig:Cluster_LJ} shows the results of a series of geometry optimisations, comparing the minimum energies, starting the optimisation from known LJ global minima\cite{hoare_lj,doye1995effect,lj_freeman,lj_northby}, as well as various known Morse (both global and local) minima configurations\cite{CCDB,doye2004global,morse_locatelli,mores_marques}, at each cluster size. For the Morse structures we only show the energy of the configuration which resulted in the lowest energy for the two tested interaction ranges. 
Due to the very short-range nature of the interaction for $r_c=1.2\sigma$, only the nearest neighbours contribute to the energy.
We found that in this case clusters typically favour the structure of those Morse global minima that had been identified for shorter attractive ranges\cite{doye1995effect}, often resulting in considerably lower energies than that of the known LJ global minimum configurations (see Fig.~\ref{fig:Cluster_LJ}). 
However, for $r_c=2.0\sigma$ the energy difference is significantly smaller, with global minimum of WRDF clusters identical to that of the LJ clusters' for the majority of cluster sizes. 

\begin{figure}[hbt]
\begin{center}
    \includegraphics[width=6.0cm,angle=90]{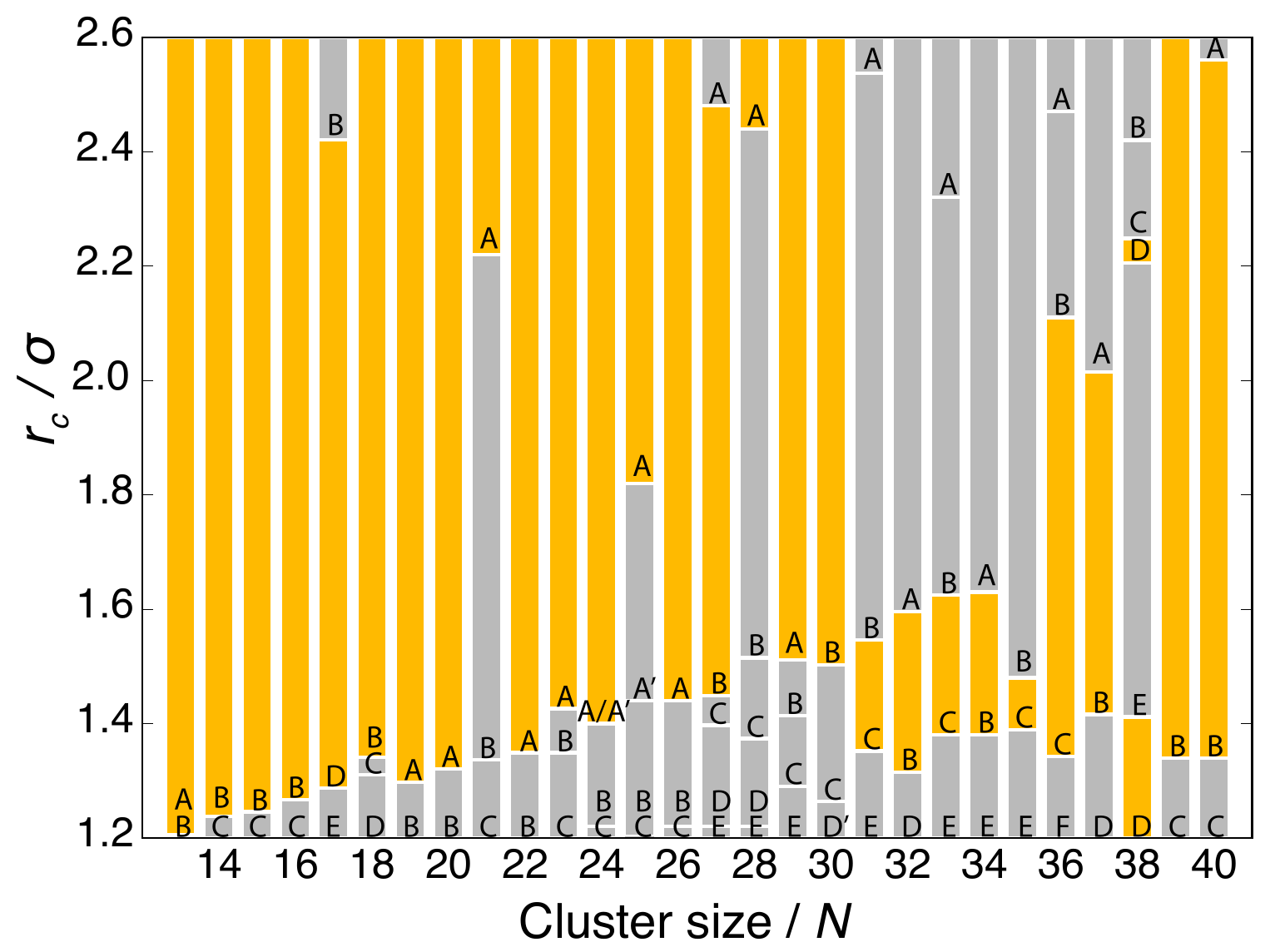}
\end{center}
\vspace{-20pt}
\caption {Identity of the global minimum cluster of WRDF particles, as a function of interaction range, $r_c$ and the number of particles, $N$. Structures are labelled from A to F, as in the database of the known Morse cluster minima.\cite{CCDB} For each cluster size, the structure that is identical to the global minimum of a Lenard-Jones cluster is shown in yellow.} 
\label{fig:Cluster_PhaseD}
\end{figure}

While the global minimum structure of small LJ clusters does not depend on the choice of truncation distance, the global minimum of the Morse potential can vary significantly as the range of the potential is changed.\cite{doye1995effect,doye2004global,morse_locatelli,mores_marques} We found that the WRDF potential behaves similarly, with the energetic order of possible minima configurations being very sensitive to the $r_c$ parameter.
Figure~\ref{fig:Cluster_PhaseD} shows the global minima structures of the WRDF clusters in the range of $N=13-40$, as a function of the $r_c$ parameter. The vertical bands represent the stability region of each structure, with the labelling following the one used in the The Cambridge Energy Landscape Database\cite{CCDB}: structure A is the Morse cluster identified for the longest range Morse potential, and consecutive letters represent structures identified for shorter ranges. It is notable that we can see a similar order of minima in our study as well.
In case of $N=13$, the cluster structure is highly symmetrical, with twelve particles surrounding the central atom. There are two competing arrangements that achieve this, one with $D_{5h}$ symmetry, the known global minimum for very short range Morse interaction (structure B), and the other with $I_h$ symmetry for longer range Morse and LJ potentials (structure B). With the WRDF potential, the $D_{5h}$ structure is the most stable for very short interaction ranges of $r_c < 1.2059 \sigma$, while the cluster favours the well-known icosahedral structure for longer cutoff values. 
A similar behaviour is observed for $N=14$, where below $r_c = 1.238 \sigma$ the Morse minimum structure with $C_{2v}$ symmetry is the global minimum (structure C), while above it is the known $C_{3v}$ structure of LJ$_{13}$ (structure B).
The truncation distance where the change in the global minimum structure happens then gradually increases with systems size, it is $r_c = 1.246\sigma$ for $N=15$ and $r_c = 1.267\sigma$ for $N=16$.
For $N=17$, a further change is observed at $r_c = 2.421\sigma$ to the $C_{3v}$ structure (structure B), another known minimum observed in case of the Morse potential, however, the known Morse structure C does not appear to be a global minimum for the 17-particle WRDF cluster. 
While the trend generally continues, there are other cluster sizes where a known Morse structure is ``skipped'' and does not become a global minimum in the expected interaction range, e.g. structure D for 31 or structure C for 32 particles.

\begin{figure}[hbt]
\begin{center}
\includegraphics[width=6.0cm,angle=90]{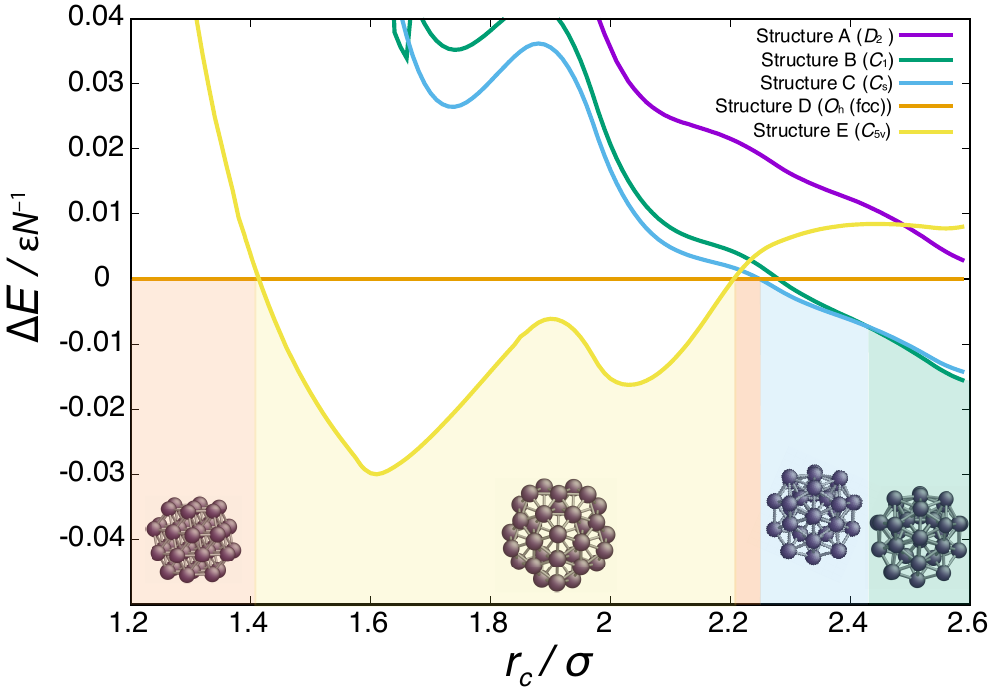}
\end{center}
\vspace{-20pt}
\caption {Comparison of minimum energies of 38 WRDF particles, arranged in different structures. Structures are labelled from A to E, as in the database of the known Morse cluster minima.\cite{CCDB} Energy difference is computed with respect to Structure D, the global minimum of LJ$_{38}$. Shaded areas highlight the interaction range where the given structure is the ground state.} 
\label{fig:Cluster38}
\end{figure}

As the size of the cluster increases, the number of competing minima increases as well. We highlight the case of $N=38$, with the energy of the optimised geometries shown in Figure~\ref{fig:Cluster38}. For $r_c<1.413\sigma$ the highly symmetric $O_h$ structure with particles in fcc arrangement is the global minimum, the same structure as for short range Morse potentials and the Lennard-Jones model. In the range $1.413\leq r_c < 2.206 \sigma$ the $C_{5v}$ icosahedral configuration has the lowest energy - this is the known global minimum for mid-range Morse potentials and the second lowest energy structure for LJ$_{38}$. Interestingly, there is a short interaction range where the $O_h$ structure becomes the global minimum again, then at $r_c=2.249\sigma$ the structure with $C_s$ symmetry becomes the most stable - this is the known third lowest energy local minimum configuration for LJ$_{38}$. However, this is also overtaken in stability by a $C_1$ structure, for $r_c>2.438\sigma$. 

\section{Conclusion}

We revisited the WRDF pair potential and evaluated its phase properties both in the bulk phase and for small cluster sizes, comparing its behaviour to that of the Lennard-Jones model.
We performed nested sampling simulations in the interaction range $r_c=1.2\sigma-2.2\sigma$ to evaluate the pressure-temperature phase diagram. At short interaction ranges the model resembles soft-matter behaviour and has a single fluid phase, with the solid-liquid-gas triple point appearing in the range $r_c=1.4\sigma-1.6\sigma$. 
Our simulations showed that the enthalpy, entropy and volume change of melting all strongly depend on the interaction change, while the melting temperature depends only weakly on the cutoff. The boiling temperature appears to be more sensitive, with the critical pressure and temperature increasing significantly for longer interaction ranges. 
The stable solid phase has a close packed structure in the entire studied range, but due to the small energy difference between different stacking variants, we observe both the face-centred-cubic and hexagonal-close-packed structures below the melting line. 
However, the ground state structure can be clearly determined, and similarly to other spherically symmetric pair potentials with non continuous higher order derivatives, we can see that the bulk-phase global minimum stacking depends on both the pressure and the interaction range, with more exotic stacking variants becoming ground state structures for $r_c>2.5\sigma$.
Structure optimisation of potential cluster structures in the size range $N=10-60$ reveal a behaviour similar to that of the Morse-clusters, with the lowest energy structure being sensitive to the truncation distance, and the clusters adapting the known Lennard-Jones global minimum structures at larger values of $r_c$ typically. These findings also confirm that changing the truncation distance of the WRDF potential changes not only the interaction range, but a broad range of thermodynamic and structural properties as well. 

\section*{Data availability}
The data supporting this study's findings are publicly available at the University of Warwick's WRAP service at \url{https://wrap.warwick.ac.uk/183723/}. The open-source package \verb|pymatnest| is freely available on GitHub at \url{https://github.com/libAtoms/pymatnest}.


\begin{acknowledgements}

The authors thank David Wales for valuable discussions regarding the minimum energy cluster structures and basin hopping calculations. 
O.A. acknowledges funding from the EPSRC Centre for Doctoral Training in Modelling of Heterogeneous Systems (EP/S022848/1).
L.B.P. acknowledges support from the EPSRC through the individual Early Career Fellowship (EP/T000163/1).
Computing facilities were provided by the Scientific Computing Research Technology Platform of the University of Warwick. 

\end{acknowledgements}



\bibliography{WangLJ}

\end{document}